\documentclass[11pt,graphicx,subfigure,axodraw]{article}
\usepackage{amsfonts}
\usepackage{amssymb}
\usepackage{amsmath,amsfonts,amssymb}
\usepackage{epstopdf}
\usepackage[section]{placeins}
\usepackage{hyperref}
\usepackage{cleveref}
\crefname{equation}{Eq.}{Eqs.}
\crefname{figure}{Fig.}{Figs.}
\crefname{table}{Table}{Tables}
\crefname{section}{Section}{Sections}

\usepackage[usenames,dvipsnames]{color}
\usepackage{graphicx}
\usepackage{subfigure}
\usepackage{float}
\def\rmuu{\gamma^{\mu}}
\def\rmud{\gamma_{\mu}}
\def\PL{{1-\gamma_5\over 2}}
\def\PR{{1+\gamma_5\over 2}}
\def\sinW2{\sin^2\theta_W}
\def\AEM{\alpha_{EM}}
\def\mul{M_{\tilde{u} L}^2}
\def\mur{M_{\tilde{u} R}^2}
\def\mdl{M_{\tilde{d} L}^2}
\def\mdr{M_{\tilde{d} R}^2}
\def\mz2{M_{z}^2}
\def\c2b{\cos 2\beta}
\def\au{A_u}
\def\ad{A_d}
\def\cob{\cot \beta}
\def\v#1{v_#1}
\def\tb{\tan\beta}
\def\epem{$e^+e^-$}
\def\KK{$K^0$-$\overline{K^0}$}
\def\wi{\omega_i}
\def\xj{\chi_j}
\def\Wmu{W_\mu}
\def\Wnu{W_\nu}
\def\m#1{{\tilde m}_#1}
\def\mH{m_H}
\def\mw#1{{\tilde m}_{\omega #1}}
\def\mx#1{{\tilde m}_{\chi^{0}_#1}}
\def\mc#1{{\tilde m}_{\chi^{+}_#1}}
\def\mwi{{\tilde m}_{\omega i}}
\def\mxi{{\tilde m}_{\chi^{0}_i}}
\def\mci{{\tilde m}_{\chi^{+}_i}}

\def\ch{{\tilde\chi^{+}_1}}
\def\c2{{\tilde\chi^{+}_2}}

\def\tt{{\tilde\theta}}

\def\tp{{\tilde\phi}}

\def\mz{M_z}
\def\sw{\sin\theta_W}
\def\cw{\cos\theta_W}
\def\cb{\cos\beta}
\def\sb{\sin\beta}
\def\rwi{r_{\omega i}}
\def\rxj{r_{\chi j}}
\def\rfp{r_f'}
\def\Kik{K_{ik}}
\def\Fq2{F_{2}(q^2)}
\def\f{\({\cal F}\)}
\def\d1{{\f(\tilde c;\tilde s;\tilde W)+ \f(\tilde c;\tilde \mu;\tilde W)}}
\def\tw{\tan\theta_W}
\def\sec2w{sec^2\theta_W}
\begin{document}
%This is dvips(k) 5.86 Cobegin{document}
\baselineskip 18pt
%t
\def\today{\ifcase\month\or
 January\or February\or March\or April\or May\or June\or
 July\or August\or September\or October\or November\or December\fi
 \space\number\day, \number\year}
\def\thebibliography#1{\section*{References\markboth
 {References}{References}}\list
 {[\arabic{enumi}]}{\settowidth\labelwidth{[#1]}
 \leftmargin\labelwidth
 \advance\leftmargin\labelsep
 \usecounter{enumi}}
 \def\newblock{\hskip .11em plus .33em minus .07em}
 \sloppy
 \sfcode`\.=1000\relax}
\let\endthebibliography=\endlist
\def\lsim{\ ^<\llap{$_\sim$}\ }
\def\gsim{\ ^>\llap{$_\sim$}\ }
\def\r2{\sqrt 2}
\def\beq{\begin{equation}}
\def\eeq{\end{equation}}
\def\beqn{\begin{eqnarray}}
\def\eeqn{\end{eqnarray}}
\def\rmuu{\gamma^{\mu}}
\def\rmud{\gamma_{\mu}}
\def\PL{{1-\gamma_5\over 2}}
\def\PR{{1+\gamma_5\over 2}}
\def\sinW2{\sin^2\theta_W}
\def\AEM{\alpha_{EM}}
\def\mul{M_{\tilde{u} L}^2}
\def\mur{M_{\tilde{u} R}^2}
\def\mdl{M_{\tilde{d} L}^2}
\def\mdr{M_{\tilde{d} R}^2}
\def\mz2{M_{z}^2}
\def\c2b{\cos 2\beta}
\def\au{A_u}
\def\ad{A_d}
\def\cob{\cot \beta}
\def\v#1{v_#1}
\def\tb{\tan\beta}
\def\epem{$e^+e^-$}
\def\KK{$K^0$-$\bar{K^0}$}
\def\wi{\omega_i}
\def\xj{\chi_j}
\def\Wmu{W_\mu}
\def\Wnu{W_\nu}
\def\m#1{{\tilde m}_#1}
\def\mH{m_H}
\def\mw#1{{\tilde m}_{\omega #1}}
\def\mx#1{{\tilde m}_{\chi^{0}_#1}}
\def\mc#1{{\tilde m}_{\chi^{+}_#1}}
\def\mwi{{\tilde m}_{\omega i}}
\def\mxi{{\tilde m}_{\chi^{0}_i}}
\def\mci{{\tilde m}_{\chi^{+}_i}}
\def\mz{M_z}
\def\sw{\sin\theta_W}
\def\cw{\cos\theta_W}
\def\cb{\cos\beta}
\def\sb{\sin\beta}
\def\rwi{r_{\omega i}}
\def\rxj{r_{\chi j}}
\def\rfp{r_f'}
\def\Kik{K_{ik}}
\def\Fq2{F_{2}(q^2)}
\def\f{\({\cal F}\)}
\def\d1{{\f(\tilde c;\tilde s;\tilde W)+ \f(\tilde c;\tilde \mu;\tilde W)}}
%%%%%%%%%%%%%%%%%%%%%%%%%%%%%%%%%%
\def\tw{\tan\theta_W}
\def\sec2w{sec^2\theta_W}
%%%%%%%%%%%%%%%%%%%%%%%%%%%%%%%%%%
\def\ch{{\tilde\chi^{+}_1}}
\def\c2{{\tilde\chi^{+}_2}}

\def\tt{{\tilde\theta}}

\def\tp{{\tilde\phi}}

\def\mz{M_z}
\def\sw{\sin\theta_W}
\def\cw{\cos\theta_W}
\def\cb{\cos\beta}
\def\sb{\sin\beta}
\def\rwi{r_{\omega i}}
\def\rxj{r_{\chi j}}
\def\rfp{r_f'}
\def\Kik{K_{ik}}
\def\Fq2{F_{2}(q^2)}
\def\f{\({\cal F}\)}
\def\d1{{\f(\tilde c;\tilde s;\tilde W)+ \f(\tilde c;\tilde \mu;\tilde W)}}

\def\b{$\cal{B}(\tau\to\mu \gamma)$~}

%%%%%%%%%%%%%%%%%%%%%%%%%%%%%%%%%%
\def\tw{\tan\theta_W}
\def\sec2w{sec^2\theta_W}
%%%%%%%%%%%%%%%%%%%%%%%%%%%%%%%%%%

\begin{titlepage}
\begin{center}
{\large {\bf Large
Neutrino Magnetic Dipole Moments  in MSSM Extensions}}\\
\vskip 0.5 true cm
Amin Aboubrahim$^{b}$\footnote{Email: amin.b@bau.edu.lb}, Tarek Ibrahim$^{a,b}$\footnote{Email: 
t.ibrahim@bau.edu.lb}, Ahmad Itani$^{b}$\footnote{Email: a.itanis@bau.edu.lb},
  and Pran Nath$^{c}$\footnote{Emal: nath@neu.edu}
\vskip 0.5 true cm
\end{center}

\noindent
{a. Department of  Physics, Faculty of Science,
University of Alexandria, Alexandria 21511, Egypt\footnote{Permanent address.} }\\
{b. Department of Physics, Faculty of Sciences, Beirut Arab University,
Beirut 11 - 5020, Lebanon\footnote{Current address.}} \\
{c. Department of Physics, Northeastern University, Boston, MA 02115-5000, USA} \\

\vskip 0.5 true cm

\centerline{\bf Abstract}
An analysis of the Dirac neutrino magnetic moment with standard model interactions gives
 $\mu_\nu\sim 3 \times 10^{-19} \mu_B (m_\nu/1 eV)$. The observation of
 a significantly larger magnetic moment will provide a clear signal of  new physics beyond the
 standard model. The current experimental limits on the neutrino magnetic moments are
 orders of magnitude larger  than the prediction with the standard model interactions 
 and thus its test appears out of reach.
  Here we give an analysis of the Dirac neutrino magnetic moments within the framework of a
  minimal supersymmetric standard model extension with a  vectorlike  lepton generation. 
  Specifically we compute the
 moments  arising from the exchange of W and the charged leptons in the loop, as well as from the exchange of
  charginos, charged sleptons and charged mirror sleptons.  It is shown that the neutrino moment
in this case  can be several orders of magnitude larger than the one with standard model like interactions,
 lying close to and below  the current experimental upper limits and should be accessible in improved
  future experiment. A correlated prediction of the heaviest neutrino lifetimes from radiative decays to
  the lighter neutrinos via exchange  of charginos and sleptons in the loops is also made. The predicted lifetimes 
  are several orders of magnitude smaller than the one with the standard model interactions and 
  also lie close to  the current experimental limits from analyses using the cosmic background neutrino data.
   \\
 \noindent
{\scriptsize
}

\medskip

\end{titlepage}
\section{Introduction \label{sec1}}
A Dirac neutrino with standard model interactions has a magnetic moment which is given by
~ \cite{shrock-fujikawa}
\beq
 \mu_{\nu_i}= \frac{3 m_e G_F}{4 \sqrt 2 \pi^2} m_{\nu_i} \simeq 3\times 10^{-19} \left(\frac{m_{\nu_i}}{eV}\right) \mu_B\ , 
 \label{1.1}
\eeq
where $\mu_B=e/2m_e$ is the Bohr Magneton (for related works see ~\cite{earlywork}). The observation of a neutrino magnetic moment 
{larger than that of \cref{1.1}}
would be a clear sign of the presence of new physics beyond  the standard model. Thus a determination  of the neutrino magnetic
moment is of great significance in the search for {physics}  beyond the standard model. There are a variety
of experimental searches which we discuss below. Thus the
Borexino experiment\cite{borexino} gives an upper bound of

\beq
\mu_\nu \leq 5.4 \times 10^{-11} \mu_B~~90\% ~{\rm CL}\ . 
\label{1.2}
\eeq
which improves the previous limit of $8.4\times 10^{-11} \mu_B$ found in   \cite{borexino-1}.
Since Borexino  explores solar neutrinos, the magnetic moment measured by Borexino  is a linear combination
of the magnetic {moments} of the three neutrino flavors.   Separately the limits on $e$, $\mu$ and $\tau$
 neutrinos are
\begin{gather}
\mu_{\nu_e} < 5.8 \times 10^{-11} \mu_B, \\
\mu_{\nu_\mu} < 1.5 \times 10^{-10} \mu_B,\\
\mu_{\nu_\tau} < 1.9 \times 10^{-10} \mu_B.
\label{1.2a}
\end{gather}
 In reactor experiments the constraint on the neutrino magnetic moment depends on the
 flavor of the initial neutrino such as in $\nu_i-e$ scattering. Thus in such experiments
 the constraint on one flavor can be gotten\footnote{In $e^+e^-$ annihilation process $e^+e^-\to \nu\bar \nu \gamma$
 constraints on the neutrino magnetic moments can also be obtained but such constraints are
 relatively weak~\cite{Beringer:1900zz}.}.  
 The TEXONO Collaboration
gives an upper limit of~\cite{Wong:2006nx}
\beq
\mu_{\nu_e} <  7.4\times 10^{-11} \mu_B,~~~~~90\% ~{\rm CL}.
\label{1.2b}
\eeq

The GEMMA experiment~\cite{Beda:2012zz} (For previous limits see ~\cite{Beda:2009kx}) gives an upper limit of  

\beq
\mu_{\nu_e} < 2.9 \times 10^{-11} \mu_B, ~~~~~~90\%~{\rm CL}\ . 
\label{1.2c}
\eeq
A more stringent limit comes from a study of the red giants at the time of helium flash, and the analysis of ~\cite{Raffelt:1990pj}
gives a constraint on the neutrino dipole moment of

\beq
\mu_\nu < 3\times 10^{-12} \mu_B\ , 
\label{1.4}
\eeq
(which is essentially a limit on the magnetic moment { $\mu_{\nu_e}$})
while the Particle Data Group~\cite{Beringer:1900zz} gives a more comprehensive list of
neutrino magnetic moment limits most of which, however, are not competitive with the
limits in \cref{1.2}-\cref{1.4}.
Recently the authors of ~\cite{Bell:2005kz} have  derived an upper bound on
the Dirac neutrino magnetic moment within a low energy effective theory and obtain a
limit of $\mu_\nu \leq 10^{-14} \mu_B$.  The derivation assumes there be no fine-tuning of
the coefficients of the  operators in the effective theory.
[The literature on neutrino magnetic moments
is extensive. For some recent works see
~\cite{Balantekin:2006sw,Kouzakov:2011mt,Geng:2012jm} and
for recent reviews see~\cite{Wong:2005pa,Broggini:2012df,Giunti:2008ve}].
In any case one finds that all of the limits above whether experimental or theory are  several orders of
magnitude larger than the estimate of \cref{1.1}.\\

In this work we carry out an analysis of the neutrino magnetic moments in an extension of MSSM which includes
a vectorlike  leptonic generation which leads to significantly larger neutrino magnetic moments than
given by \cref{1.1}.
Vector like generations arise in many GUT and string models~\cite{guts}
 and have been
discussed in the literature quite frequently~
\cite{Aboubrahim:2013gfa,Ibrahim:2008gg,Ibrahim:2010va,Ibrahim:2011im,Ibrahim:2010hv,Ibrahim:2012ds,Ibrahim:2009uv,Babu:2008ge,Liu:2009cc,Martin:2009bg,Hewett:2012ns}.\\

The outline of the rest of the paper is as follows: In \cref{sec2} we describe briefly the framework of the model which
is an extension of the minimal supersymmetric standard model (MSSM)
 including a vector leptonic multiplet. Here we define the basic interactions that mix the vector
like generation with the regular three generations of leptons {and sleptons}. These mixings arise via the superpotential couplings
as well as via soft breaking.
In \cref{sec3} we present the interactions of the neutrinos with W bosons, leptons, and their mirrors and in 
  \cref{sec4} we give the interactions of the neutrinos with charginos, sleptons and mirror sleptons.
   Using these interactions a full one loop analysis of the neutrino magnetic moments is  carried out in \cref{sec5}
   where the contributions  from the exchange of the W and the charged leptons arise via loops shown in \cref{fig1a} 
and the contributions from the exchange of charginos, sleptons and mirror sleptons arise via loops shown
in \cref{fig1b}. 
 In \cref{sec6} we give a
numerical analysis of the size of effects as a result of the mixing of the vectorlike  generation with the three
regular generations of leptons {and sleptons}. The analysis shows that neutrino magnetic moments as large as $({10^{-10}}-10^{-14}) \mu_B$ can be
gotten in models discussed in \cref{sec5} (For large neutrino magnetic moments arising from
large extra dimensions see ~\cite{  McLaughlin:1999br,Mohapatra:2004ce}).  In this section 
we also give an evaluation of the $\nu_3$ lifetime on which experimental lower limits exist from 
radiative decays of the cosmic background neutrinos. 
 Conclusions are given in \cref{sec7}. Further details of the model are given
in \cref{sec8}.

\section{MSSM Extension with a vector leptonic multiplet\label{sec2}}

Vector like multiplets arise in a variety of unified models~\cite{guts} some of which could be low lying. Here we simply assume
the existence of one low lying leptonic vector multiplet which is anomaly free  in addition to the MSSM spectrum.  Before proceeding further
it is useful to record the  quantum numbers of the leptonic matter content of this extended MSSM spectrum under $SU(3)_C\times SU(2)_L\times U(1)_Y$.
Thus  under $SU(3)_C\times SU(2)_L\times U(1)_Y$ the leptons of the
 three generations transform as follows

 \begin{align}
\psi_{iL}\equiv
 \left(\begin{matrix} \nu_{i L}\cr
 ~{l}_{iL}  \end{matrix} \right) &&
(1,2,- \frac{1}{2}),\nonumber\\
 l^c_{iL}&&(1,1,1),\nonumber\\
 \nu^c_{i L}&&(1,1,0)\ .
 % ~i=1,2,3
\label{2}
\end{align}
where the last entry on the right hand side column  is the value of the hypercharge
 $Y$ defined so that $Q=T_3+ Y$.  These leptons have $V-A$ interactions.
We can now add a vectorlike  multiplet where we have a fourth family of leptons with $V-A$ interactions
whose transformations can be gotten from Eq.(\ref{2}) by letting i run from 1-4.
A vectorlike  lepton multiplet also has  mirrors and so we consider these mirror
leptons which have $V+A$ interactions. Its quantum numbers are given by

\begin{align}
\chi^c\equiv
 \left(\begin{matrix} E_{ L}^c\cr
 N_L^c\end{matrix}\right)&&
(1,2,\frac{1}{2}),\nonumber\\
 E_{ L} && (1,1,-1),\nonumber\\
 N_L && (1,1,0).
\label{3}
\end{align}
Interesting new physics arises when we allow mixings of the vectorlike  generation with
the three ordinary generations.  Thus the  superpotential of the model allowing for the mixings
among the three ordinary generations and the vectorlike  generation is given by

\begin{multline}
W= -\mu \epsilon_{ij} \hat H_1^i \hat H_2^j+\epsilon_{ij}  [f_{1}  \hat H_1^{i} \hat \psi_L ^{j}\hat \tau^c_L
 +f_{1}'  \hat H_2^{j} \hat \psi_L ^{i} \hat \nu^c_{\tau L}
+f_{2}  \hat H_1^{i} \hat \chi^c{^{j}}\hat N_{L}
 +f_{2}'  H_2^{j} \hat \chi^c{^{i}} \hat E_{ L}\\
+ h_{1}  H_1^{i} \hat\psi_{\mu L} ^{j}\hat\mu^c_L
 +h_{1}'  H_2^{j} \hat\psi_{\mu L} ^{i} \hat\nu^c_{\mu L}
+ h_{2}  H_1^{i} \hat\psi_{e L} ^{j}\hat e^c_L
 +h_{2}'  H_2^{j} \hat\psi_{e L} ^{i} \hat\nu^c_{e L}]\\
+ f_{3} \epsilon_{ij}  \hat\chi^c{^{i}}\hat\psi_L^{j}
 + f_{3}' \epsilon_{ij}  \hat\chi^c{^{i}}\hat\psi_{\mu L}^{j}
 + f_{4} \hat\tau^c_L \hat E_{ L}  +  f_{5} \hat\nu^c_{\tau L} \hat N_{L}
 + f_{4}' \hat\mu^c_L \hat E_{ L}  +  f_{5}' \hat\nu^c_{\mu L} \hat N_{L}\\
+ f_{3}'' \epsilon_{ij}  \hat\chi^c{^{i}}\hat\psi_{e L}^{j}
 + f_{4}'' \hat e^c_L \hat E_{ L}  +  f_{5}'' \hat\nu^c_{e L} \hat N_{L}\ , 
 \label{5}
\end{multline}
where  $\hat ~$ implies superfields. 
The mass terms for the leptons and  mirror leptons arise from the term
\beq
{\cal{L}}=-\frac{1}{2}\frac{\partial ^2 W}{\partial{A_i}\partial{A_j}}\psi_ i \psi_ j+H.c.
\label{6}
\eeq
where $\psi$ and $A$ stand for generic two-component fermion and scalar fields.
After spontaneous breaking of the electroweak symmetry, ($\langle H_1^1 \rangle=v_1/\sqrt{2} $ and $\langle H_2^2\rangle=v_2/\sqrt{2}$),
we have the following set of mass terms written in the 4-component spinor notation
so that
\beq
-{\cal L}_m= \bar\xi_R^T (M_f) \xi_L +\bar\eta_R^T(M_g) \eta_L +H.c.,
\eeq
where the basis vectors in which the mass matrix is written is given by
\begin{gather}
\bar\xi_R^T= \left(\begin{matrix}\bar \nu_{\tau R} & \bar N_R & \bar \nu_{\mu R}
&\bar \nu_{e R} \end{matrix}\right),\nonumber\\
\xi_L^T= \left(\begin{matrix} \nu_{\tau L} &  N_L &  \nu_{\mu L}
& \nu_{e L} \end{matrix}\right) \ ,\nonumber\\
\bar\eta_R^T= \left(\begin{matrix}\bar{\tau_ R} & \bar E_R & \bar{\mu_ R}
&\bar{e_ R} \end{matrix}\right),\nonumber\\
\eta_L^T= \left(\begin{matrix} {\tau_ L} &  E_L &  {\mu_ L}
& {e_ L} \end{matrix}\right) \ ,
\end{gather}
and the mass matrix $M_f$ is given by

\beqn
M_f=
 \left(\begin{matrix} f'_1 v_2/\sqrt{2} & f_5 & 0 & 0 \cr
 -f_3 & f_2 v_1/\sqrt{2} & -f_3' & -f_3'' \cr
0&f_5'&h_1' v_2/\sqrt{2} & 0 \cr
0 & f_5'' & 0 & h_2' v_2/\sqrt{2}\end{matrix} \right)\ . 
\label{7}
\eeqn
The mass matrix is not hermitian and thus one needs bi-unitary transformations to diagonalize it.
We define the bi-unitary transformation so that

\beq
D^{\nu \dagger}_R (M_f) D^\nu_L=diag(m_{\psi_1},m_{\psi_2},m_{\psi_3}, m_{\psi_4} ).
\label{7a}
\eeq
Under the bi-unitary transformations the basis vectors transform so that
\beqn
 \left(\begin{matrix} \nu_{\tau_R}\cr
 N_{ R} \cr
\nu_{\mu_R} \cr
\nu_{e_R} \end{matrix}\right)=D^{\nu}_R \left(\begin{matrix} \psi_{1_R}\cr
 \psi_{2_R}  \cr
\psi_{3_R} \cr
\psi_{4_R}\end{matrix}\right), \  \
\left(\begin{matrix} \nu_{\tau_L}\cr
 N_{ L} \cr
\nu_{\mu_L} \cr
\nu_{e_L}\end{matrix} \right)=D^{\nu}_L \left(\begin{matrix} \psi_{1_L}\cr
 \psi_{2_L} \cr
\psi_{3_L} \cr
\psi_{4_L}\end{matrix}\right) \ . 
\label{8}
\eeqn
{
In \cref{7a}
$\psi_1, \psi_2, \psi_3, \psi_4$ are the mass eigenstates for the neutrinos,
where in the limit of no mixing
we identify $\psi_1$ as the light tau neutrino, $\psi_2$ as the
heavier mass eigen state,  $\psi_3$ as the muon neutrino and $\psi_4$ as the electron neutrino.
To make contact with the normal neutrino hierarchy we relabel the states so that
\beq
\nu_1= \psi_4, \nu_2= \psi_3, \nu_3= \psi_1, \nu_4= \psi_2\ , 
\eeq
which we assume has the mass hierarchical pattern
\beqn
m_{\nu_1}< m_{\nu_2} < m_{\nu_3} < m_{\nu_4}\ . 
\label{8.a}
\eeqn
We will carry out the analytical analysis in the $\psi_i$ notation but the numerical analysis
will be carried out in the $\nu_i$ notation to make direct contact with data.
A similar analysis goes to the lepton mass matrix $M_\ell$ where
\beqn
M_\ell=
 \left(\begin{matrix} f_1 v_1/\sqrt{2} & f_4 & 0 & 0 \cr
 f_3 & f'_2 v_2/\sqrt{2} & f_3' & f_3'' \cr
0&f_4'&h_1 v_1/\sqrt{2} & 0 \cr
0 & f_4'' & 0 & h_2 v_1/\sqrt{2}\end{matrix} \right)\ . 
\label{7}
\eeqn

  Next we  consider  the mixing of the charged sleptons and the charged mirror sleptons.
The mass squared  matrix of the slepton - mirror slepton comes from three sources:  the F term, the
D term of the potential and the soft susy breaking terms.
Using the  superpotential of \cref{5} the mass terms arising from it
after the breaking of  the electroweak symmetry are given by
the Lagrangian
\beq
{\cal L}= {\cal L}_F +{\cal L}_D + {\cal L}_{\rm soft}\ , 
\eeq
where   $ {\cal L}_F$ is deduced from \cref{5} and is given in the Appendix, while the ${\cal L}_D$ is given by
\begin{multline}
-{\cal L}_D=\frac{1}{2} m^2_Z \cos^2\theta_W \cos 2\beta \{\tilde \nu_{\tau L} \tilde \nu^*_{\tau L} -\tilde \tau_L \tilde \tau^*_L
+\tilde \nu_{\mu L} \tilde \nu^*_{\mu L} -\tilde \mu_L \tilde \mu^*_L
+\tilde \nu_{e L} \tilde \nu^*_{e L} -\tilde e_L \tilde e^*_L\\
+\tilde E_R \tilde E^*_R -\tilde N_R \tilde N^*_R\}
+\frac{1}{2} m^2_Z \sin^2\theta_W \cos 2\beta \{\tilde \nu_{\tau L} \tilde \nu^*_{\tau L}
 +\tilde \tau_L \tilde \tau^*_L
+\tilde \nu_{\mu L} \tilde \nu^*_{\mu L} +\tilde \mu_L \tilde \mu^*_L\\
+\tilde \nu_{e L} \tilde \nu^*_{e L} +\tilde e_L \tilde e^*_L
-\tilde E_R \tilde E^*_R -\tilde N_R \tilde N^*_R +2 \tilde E_L \tilde E^*_L -2 \tilde \tau_R \tilde \tau^*_R
-2 \tilde \mu_R \tilde \mu^*_R -2 \tilde e_R \tilde e^*_R
\}.
\label{12}
\end{multline}
For ${\cal L}_{\rm soft}$ we assume the following form
\begin{multline}
-{\cal L}_{soft}=\tilde M^2_{\tau L} \tilde \psi^{i*}_{\tau L} \tilde \psi^i_{\tau L}
+\tilde M^2_{\chi} \tilde \chi^{ci*} \tilde \chi^{ci}
+\tilde M^2_{\mu L} \tilde \psi^{i*}_{\mu L} \tilde \psi^i_{\mu L}
+\tilde M^2_{e L} \tilde \psi^{i*}_{e L} \tilde \psi^i_{e L}
+\tilde M^2_{\nu_\tau} \tilde \nu^{c*}_{\tau L} \tilde \nu^c_{\tau L}
 +\tilde M^2_{\nu_\mu} \tilde \nu^{c*}_{\mu L} \tilde \nu^c_{\mu L}\\
+\tilde M^2_{\nu_e} \tilde \nu^{c*}_{e L} \tilde \nu^c_{e L}
+\tilde M^2_{\tau} \tilde \tau^{c*}_L \tilde \tau^c_L
+\tilde M^2_{\mu} \tilde \mu^{c*}_L \tilde \mu^c_L
+\tilde M^2_{e} \tilde e^{c*}_L \tilde e^c_L
+\tilde M^2_E \tilde E^*_L \tilde E_L
 + \tilde M^2_N \tilde N^*_L \tilde N_L \\
+\epsilon_{ij} \{f_1 A_{\tau} H^i_1 \tilde \psi^j_{\tau L} \tilde \tau^c_L
-f'_1 A_{\nu_\tau} H^i_2 \tilde \psi ^j_{\tau L} \tilde \nu^c_{\tau L}
+h_1 A_{\mu} H^i_1 \tilde \psi^j_{\mu L} \tilde \mu^c_L
-h'_1 A_{\nu_\mu} H^i_2 \tilde \psi ^j_{\mu L} \tilde \nu^c_{\mu L}\\
+h_2 A_{e} H^i_1 \tilde \psi^j_{e L} \tilde e^c_L
-h'_2 A_{\nu_e} H^i_2 \tilde \psi ^j_{e L} \tilde \nu^c_{e L}
+f_2 A_N H^i_1 \tilde \chi^{cj} \tilde N_L
-f'_2 A_E H^i_2 \tilde \chi^{cj} \tilde E_L +H.c.\}\ . 
\label{13}
\end{multline}

\begin{figure}[h]
\begin{center}
{\rotatebox{0}{\resizebox*{10cm}{!}{\includegraphics{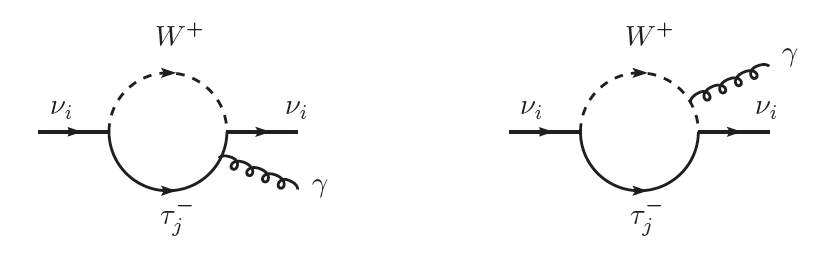}}\hglue5mm}}
\caption{The diagrams that contribute to neutrino magnetic dipole moment via
  exchange of the $W$ boson and of the leptons and of the  mirror leptons
  where the photon is either emitted by the $W$ boson {(right)} or by the lepton or by the mirror lepton
   {(left)} inside the loop.} \label{fig1a}
\end{center}
\end{figure}

\begin{figure}[h]
\begin{center}
{\rotatebox{0}{\resizebox*{10cm}{!}{\includegraphics{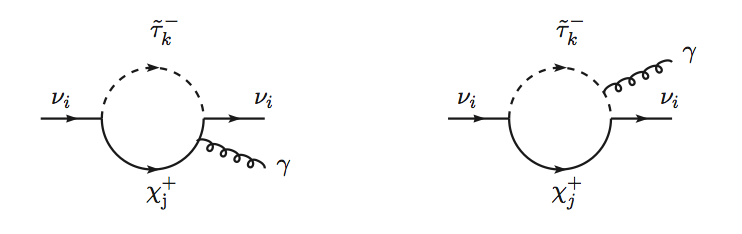}}\hglue5mm}}
\caption{The diagrams that contribute to neutrino magnetic dipole moment via
 supersymmetric loops involving the exchange of charginos and of sleptons and mirror sleptons
  where the photon is either emitted by the chargino (left) or by the slepton or by the mirror slepton (right) inside the loop.} \label{fig1b}
\end{center}
\end{figure}

\section{Interactions of W boson, leptons, mirrors and neutrinos\label{sec3}}
The magnetic dipole moments of the neutrinos arise from the loop diagrams of \cref{fig1a} and \cref{fig1b}. We now write the charged current interaction in the leptonic sector for the three generations and for the
mirror generation with the W boson,
\beqn
-{\cal L}_{CC}= \frac{g}{2 \sqrt{2}} W^{\dagger}_{\rho} \{ \bar{\nu}_{\tau} \gamma^{\rho} (1-\gamma_5) \tau
+ \bar{\nu}_{\rho} \gamma^{\mu} (1-\gamma_5) \mu
+ \bar{\nu}_{e} \gamma^{\rho} (1-\gamma_5) e
+ \bar{N} \gamma^{\rho} (1+\gamma_5) E\}+H.c.
\eeqn
Using Eq.{\ref{8}}, and its counterpart in the lepton sector, one can write the charged current interactions in the mass diagonal basis as
\begin{multline}
-{\cal L}_{CC}= \frac{g}{2 \sqrt{2}} W^{\dagger}_{\rho}\sum_{i=1}^{4}\sum_{j=1}^{4}   \bar{\psi}_{i} \gamma^{\rho} \{
(D^{\nu *}_{L 1i} D^{\tau}_{L 1j} + D^{\nu *}_{L 3i} D^{\tau}_{L 3j}\\
+D^{\nu *}_{L 4i} D^{\tau}_{L 4j}) (1-\gamma_5)
+(D^{\nu *}_{R 2i} D^{\tau}_{R 2j})  (1+\gamma_5) \} \tau_j\  + H.c. . 
\end{multline}

\section{Interactions of charginos, sleptons and neutrinos \label{sec4}}
The magnetic dipole moments of the neutrinos arise from loop diagrams of  Fig.2.
 The relevant part of  the Lagrangian that generates this contribution is given by
\beqn
-{\cal{L}}_{\nu-\tilde{\tau}-\chi^+}=
\sum_{i =1}^4\sum_{j=1}^2\sum_{k=1}^8
\bar{\psi}_{i}[C^L_{i j  k} P_L+
 C^R_{i j k} P_R]
\tilde{\chi}^+_j \tilde{\tau}_k +H.c.
\label{16}
\eeqn
where
\begin{multline}
C^L_{i j k}=-f'_1  V^*_{j2} D^{\nu *}_{R _{1 i}}\tilde{D}^{\tau}_{1 k}
-f'_2  V^*_{j 2} D^{\nu *}_{R _{2 i}}\tilde{D}^{\tau}_{2 k}
+g  V^*_{j1} D^{\nu *}_{R _{2 i}}\tilde{D}^{\tau}_{4 k}
-h'_1  V^*_{j2} D^{\nu *}_{R _{3 i}}\tilde{D}^{\tau}_{5 k}
-h'_2  V^*_{j2} D^{\nu *}_{R _{4 i}}\tilde{D}^{\tau}_{7 k},
\\
C^R_{i j k}=-f_1  U_{j2} D^{\nu *}_{L _{1 i}}\tilde{D}^{\tau}_{3 k}
-h_1  U_{j2} D^{\nu *}_{L _{3 i}}\tilde{D}^{\tau}_{6 k}
+g  U_{j1} D^{\nu *}_{L _{1 i}}\tilde{D}^{\tau}_{1 k}
+g  U_{j1} D^{\nu *}_{L _{4 i}}\tilde{D}^{\tau}_{7 k}\\
-h_2  U_{j2} D^{\nu *}_{L _{4 i}}\tilde{D}^{\tau}_{8 k}
-f_2  U_{j2} D^{\nu *}_{L _{2 i}}\tilde{D}^{\tau}_{4 k},
\label{17}
\end{multline}
where $\tilde{D}^{\tau}$ is the diagonalizing matrix of the scalar mass squared  matrix
for the scalar leptons as defined
in the Appendix.
In Eq.(\ref{17})
 $U$ and $V$ are the matrices  that  diagonalize the chargino mass matrix $M_C$
  so that
\beq
U^* M_C V^{-1}= diag (m_{\tilde{\chi_1}}^+,m_{\tilde{\chi_2}}^+).
\label{19}
\eeq

\section{An analytical computation of the neutrino magnetic moments\label{sec5}}
The dipole moments for neutrinos are  defined by

\beq
<\nu_j(p')|J_{\alpha}^{em}|\nu_i( p ) >=   \mu^m_\nu(q^2)_{ij} \bar u_j(p')i\sigma_{\alpha\rho} q^\rho u_i ( p )+
d^e_\nu(q^2)_{ij} \bar u_j(p')i\sigma_{\alpha\rho} \gamma_5 q^\rho u_i ( p )+\cdots
\eeq
where $ \mu^m_\nu(q^2)_{ij}$ is the magnetic dipole form factor and  $ d^e_\nu(q^2)_{ij}$ is the electric
dipole form factor. We are interested in their values at zero momentum transfer, i.e., the quantities
$\mu^m_\nu(0)_{ij}$,  $ d^e_\nu(0)_{ij}$. For the case when $i=j$ these moments vanish if the
neutrino is a Majorana field, but is in general non-vanishing for a Dirac neutrino. When $i\neq j$ one has
 transition moments and they can be non-vanishing both for Dirac as well as for  Majorana neutrinos.
 The analysis given below is general in that we consider the lepton and the  neutrino mass mixings which form
   $4\times 4$  {matrices} while the slepton mass squares form an $8\times 8$ matrix. A sub case was previously considered in 
 \cite{Ibrahim:2010va} where only the tau neutrino magnetic moment was computed while the analysis given here is
 more general. \\

We now give a computation of the magnetic dipole moment arising from the loop diagrams of Fig.1 and Fig. 2.
First  the W boson loops of Fig. 1 produce the following contribution  to the magnetic moment $\mu_i$ of the neutrino  $\psi_i$  in  Bohr magneton units $\mu_B (=e/2m_e)$ so that 
\beqn
\mu^W_i=-\frac{g^2 m_e}{64 \pi^2 M^2_W} \sum_{j=1}^{4} m_{\tau_j} [|v_{ij}|^2 - |a_{ij}|^2] G_1(\frac{m_{\tau_j}}{M_W})+\nonumber\\
\frac{3 g^2 m_e m_{\psi_i}}{128 \pi^2 M^2_W}\sum_{j=1}^{4} [|v_{ij}|^2 + |a_{ij}|^2] G_2(\frac{m_{\tau_j}}{M_W})\ , 
\eeqn
where
\beqn
v_{ij}=D^{\nu *}_{L 1i} D^{\tau}_{L 1j} + D^{\nu *}_{L 3i} D^{\tau}_{L 3j}+D^{\nu *}_{L 4i} D^{\tau}_{L 4j}+D^{\nu *}_{R 2i} D^{\tau}_{R 2j}   ,\nonumber\\
a_{ij}=D^{\nu *}_{L 1i} D^{\tau}_{L 1j} + D^{\nu *}_{L 3i} D^{\tau}_{L 3j}+D^{\nu *}_{L 4i} D^{\tau}_{L 4j}-D^{\nu *}_{R 2i} D^{\tau}_{R 2j}\ . 
\eeqn
The chargino exchange loops of Fig. 2 produce a contribution  to the magnetic moments of the neutrinos in $\mu_B$ units so that 
\beqn
\mu^{\chi}_i=\frac{m_e}{16 \pi^2} \sum_{j=1}^2 \sum_{k=1}^{8} \frac{1}{m_{\chi_j}}
 [|v_{ijk}|^2 - |a_{ijk}|^2] G_3(\frac{m_{\tilde{\tau_k}}}{m_{\chi_j}})+
\frac{m_e m_{\psi_i}}{48 \pi^2} [|v_{ijk}|^2 +|a_{ijk}|^2] G_4(\frac{m_{\tilde{\tau_k}}}{m_{\chi_j}})\ , 
\eeqn
where
\beqn
v_{ijk}=\frac{1}{2} \{C^L_{ijk} + C^R_{ijk}\}\ , \nonumber\\
a_{ijk}=\frac{1}{2}\{C^L_{ijk}-C^R_{ijk}\}\ . 
\eeqn
The form factors $G_i(x)$ are given by
\beqn
G_1(x)=\frac{4-x^2}{1-x^2}+\frac{3 x^2}{(1-x^2)^2}\ln(x^2),\nonumber\\
G_2(x)=\frac{2-5x^2+x^4}{(1-x^2)^2} -\frac{2x^4}{(1-x^2)^3}\ln(x^2),\nonumber\\
G_3(x)=\frac{-2}{x^2-1}+\frac{2 x^2}{(x^2-1)^2}\ln(x^2),\nonumber\\
G_4(x)=\frac{3(1+x^2)}{(1-x^2)^2}+\frac{6 x^2}{(1-x^2)^3}\ln(x^2)\ . 
\eeqn
The form factors $G_1$ and $G_2$ arise from the non-supersymmetric loops of \cref{fig1a} 
involving the exchange of the W boson and the charged leptons while the {form} factors $G_3$ and 
$G_4$ arise from the supersymmetric loops of \cref{fig1b} from the exchange of the 
charginos and the charged sleptons. 

\section{Numerical analysis and results\label{sec6}}
In this section we give a numerical analysis for the magnetic moment of  the electron ($\mu_1$) and for the 
muon ($\mu_2$). In the analysis we will impose the constraint on the sum of the neutrino masses 
arising from the Planck Satellite experiment~\cite{Ade:2013ktc} so that 
 \beq
\sum_{i=1}^3m_{\nu_{i}}<0.85 eV \ , 
\label{6.1a}
\eeq
where we assume $\nu_i$ (i=1,2,3) to be the mass eigenstates with eigenvalues $m_{\nu_i}$ 
with the mass hierarchy given {in} \cref{8.a}.
Neutrino oscillations  constraint the neutrino mass squared differences so that~\cite{Schwetz:2008er}
\begin{gather}
\label{6.1b}
\Delta m^2_{31}\equiv m_3^2-m_1^2= 2.4^{+0.12}_{-0.11} \times 10^{-3} ~eV^2  \ , \\
\Delta m_{21}^2\equiv m_2^2- m_1^2= 7.65^{+0.23}_{-0.20} \times 10^{-5}~eV^2. 
\label{6.1c}
\end{gather}

%%%
 \begin{figure}[h]
\begin{center}
{\rotatebox{0}{\resizebox*{13cm}{!}{\includegraphics{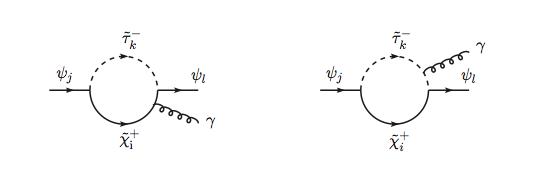}}\hglue5mm}}
\caption{Neutrino radiative decay 
$\psi_j\to \psi_l+\gamma$ via supersymmetric loops involving the charginos and the {sleptons}
 by  the emission of the photon from either the
 chargino (left) or by the {slepton} (right) inside the loop.} \label{taumugamma}
\end{center}
\end{figure}

It is also interesting to include in the analysis of the neutrino magnetic moments the 
prediction of the neutrino lifetime for the decay $\nu_3\to \nu_1\gamma, \nu_2\gamma$ (see \cref{taumugamma}).
A computation of the {neutrino} lifetime {with} the standard model {interactions} gives ~\cite{Pal:1981rm}
\beqn
 \tau_{\nu_3}^{SM}  \sim 10^{43} ~~{\rm yrs},
 \label{00.22}
  \eeqn
for a $\nu_3$ with mass 50 meV.  One may compare this with the experimental data 
from galaxy surveys within frared satellites AKARI~\cite{Matsuura:2010rb}, Spitzer~\cite{Dole:2006de} and Hershel~\cite{Berta:2010rc}  as well as the high precision
cosmic microwave background (CMB) data collected by the Far Infrared Absolute Spectrometer (FIRAS) on board the
Cosmic Background explorer (COBE)~\cite{Mirizzi:2007jd}
 for the study of radiative decays of the cosmic neutrinos\cite{Kim:2011ye} using the Cosmic Infrared Background (CIB) which gives \cite{Kim:2011ye}
 \beqn
  \tau_{\nu_3}^{exp}  \geq  10^{12} ~~{\rm yrs}\ .
  \label{6.3}
  \eeqn
  A lifetime orders of magnitude smaller than that of Eq.(\ref{00.22}) was shown to arise in the 
  analysis of \cite{Aboubrahim:2013gfa}. 
  In the analysis of $\nu_3$ lifetime  here  we will use the formulae derived in~\cite{Aboubrahim:2013gfa}.  
  As in ~\cite{Aboubrahim:2013gfa} we will see that similar size lifetimes also arise
  with the inputs used in the analysis of the neutrino magnetic dipole moments  presented here.  \\

In \cref{table1} we give the analysis of neutrino masses  for four sets of inputs.
For each set of inputs, in addition to the muon and the electron neutrino magnetic moments,
we also display the $\nu_3$ lifetime as well as the {lighter}
chargino and slepton masses that enter the loops.
The computed values of the  tau neutrino lifetime are consistent with the analysis of  \cite{Aboubrahim:2013gfa}.
 From \cref{table1} it is clear that $\mu_{2}$ can lie close to $\cal{O}$$(10^{-10})\mu_{B}$ which is  the current experimental limit. Further $\mu_{1}$  can be $\cal{O}$$(10^{-12})\mu_{B}$. Both $\mu_1$
 and $\mu_2$ are several orders in magnitude greater than that predicted by the Standard Model type interactions.

\begin{table}[H]
\centering
\begin{tabular}{ccc}
  \hline\hline
  & Neutrino Mass Eigenvalues   &$m_{\nu_{3}}=5.2\times10^{-11}$ \\
       &   (GeV)                        & $m_{\nu_{2}}=9.2\times10^{-12}$  \\
    &                          & $m_{\nu_{1}}=9.7\times10^{-13}$ \\
     \hline
(i) ~$m_{\chi^{\pm}}=256$ GeV    & $\mu_{2}$ & $1.2\times10^{-10}$ \\
 ~~~$m_{\tilde{\tau}}=162$ GeV    & $\mu_{1}$ & $2.5\times10^{-13}$ \\ 
    &  $\nu_3$ lifetime & $3.9\times10^{14}$ yrs\\
 (ii)~$m_{\chi^{\pm}}=267$ GeV    & $\mu_{2}$ & $4.6\times10^{-10}$ \\
 ~~~$m_{\tilde{\tau}}=202$ GeV        & $\mu_{1}$ & $1.3\times10^{-12}$ \\
    &$\nu_3$ lifetime & $2.5\times10^{14}$ yrs\\
(iii) ~ $m_{\chi^{\pm}}=268$ GeV    & $\mu_{2}$ & $2.2\times10^{-10}$ \\
 ~~~~$m_{\tilde{\tau}}=158$ GeV   
     & $\mu_{1}$ & $1.1\times10^{-13}$ \\
    & $\nu_3$ lifetime & $1.8\times10^{14}$ yrs\\
 (iv)~~$m_{\chi^{\pm}}=272$ GeV    & $\mu_{2}$ & $-7.6\times10^{-10}$ \\
 ~~~~$m_{\tilde{\tau}}=195$ GeV   
    & $\mu_{1}$ & $-1.3\times10^{-13}$ \\
    & $\nu_3$ lifetime & $8.8\times10^{13}$ yrs \\
    \hline\hline

\end{tabular}
\caption{An exhibition of the
numerical values for the muon neutrino magnetic moment $\mu_{2}$ and of  the electron neutrino magnetic moment $\mu_{1}$ for four sets of inputs (i)-(iv).
The common parameter for the four sets are:
   $|f_{3}|=7\times10^{-8}$, $|f_{3}'|=5\times10^{-8}$, $|f_{3}''|=8\times10^{-9}$, $|f_{4}'|=|f_{4}''|=42$, $|f_{5}|=8.11\times10^{-2}$, $|f_{5}'|=9.8\times10^{-2}$, $|f_{5}''|=4\times10^{-2}$, $m_{N}=212$, $\tan\beta=60$, $\chi_{3}=0.3$, $\chi_{3}'=0.2$, $\chi_{3}''=0.6$, $\chi_{4}=3.1$, $\chi_{4}'=0.1$, $\chi_{4}''=0.5$, $\chi_{5}=1.9$, $\chi_{5}'=0.5$ and $\chi_{5}''=0.7$. Additional parameters which are different for different sets are as follows:   
    Set (i): $m_{E}=460$, $m_{0}=300$, $A_{0}=579$, $m_{2}=320$, $\mu=300$ and $|f_{4}|=65$. Set (ii): $m_{E}=550$, $m_{0}=300$, $A_{0}=600$, $m_{2}=350$, $\mu=300$ and $|f_{4}|=65$. Set (iii): $m_{E}=550$, $m_{0}=305$, $A_{0}=650$, $m_{2}=355$, $\mu=300$ and $|f_{4}|=65$. Set (iv): $m_{E}=703$, $m_{0}=335$, $A_{0}=780$, $m_{2}=320$, $\mu=305$ and $|f_{4}|=64$.  All masses are in GeV, phases in rad and magnetic moments in units of $\mu_{B}$. The neutrino mass eigenvalues and the $\nu_3$ lifetime
    are also exhibited.   
    } \label{table1}
\end{table}

\cref{fig1} displays the neutrino magnetic moments $\mu_1$ and $\mu_2$ as a function of the soft 
 $SU(2)$ gaugino mass $m_{2}$.   The gaugino mass $m_{2}$ enters the analysis via the 
 chargino mass matrix. The analysis of \cref{fig1} is for three values of $\tan\beta$ which from 
  top to bottom are 40, 50 and 60 which correspond to the unmarked 
  solid, long -dashed and short-dashed curves.  
  We note that the largest contribution to
 the magnetic moments arise from the supersymmetric sector, i.e., from the chargino exchange
 loop diagrams while the W exchange loop diagrams make a negligible contribution. 
   Fig.~\ref{fig1} shows that $\mu_{2}$ can be $\cal{O}$$(10^{-10}\mu_{B})$  and the predicted values fall 
   below but are close to the  experimental upper limit. As for $\mu_{1}$, the predicted    
   values can reach $\sim10^{-13}\mu_{B}$ which is clearly a major enhancement on what is predicted by the Standard Model like interactions. The $\nu_3$ lifetime corresponding to the cases $\tan\beta=40,50,60$ is also exhibited by the marked curves.
   \\

\begin{figure}[H]
\begin{center}
\hfill
\subfigure[The muon neutrino magnetic moment and the tau neutrino lifetime versus $m_{2}$.]
{\includegraphics[width=8cm]{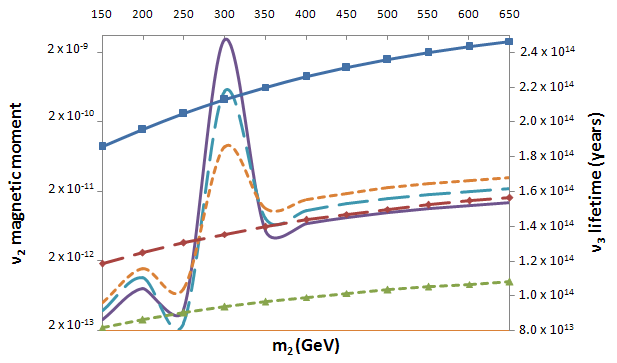}}
\hfill
\subfigure[The electron neutrino magnetic moment and the tau neutrino lifetime versus $m_{2}$.]
{\includegraphics[width=8cm]{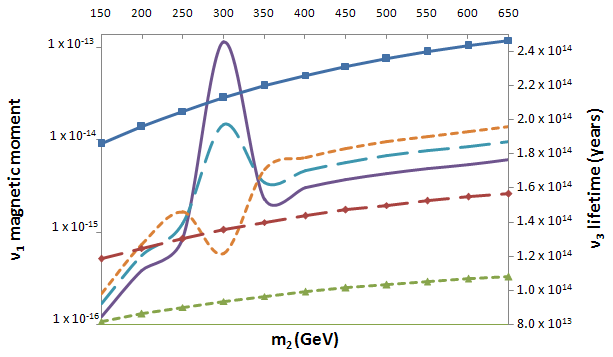}}
\hfill
\caption{A display of the neutrino magnetic moments (unmarked solid, long dashed and short dashed)
and  the $\nu_{3}$ lifetime (marked solid, long dashed and short dashed)
as a function of the gaugino mass $m_{2}$
in the range 150-650 GeV. The three  sets of curves correspond to $\tan\beta=40$ (solid curve), $\tan\beta=50$ (long dashed), and $\tan\beta=60$ (short dashed).  Other parameters have the values $\mu=100$, $|f_{3}|=7\times10^{-8}$, $|f_{3}'|=5\times10^{-8}$, $|f_{3}''|=8\times10^{-9}$, $|f_{4}|=57$, $|f_{4}'|=42$, $|f_{4}''|=42$, $|f_{5}|=8.11\times10^{-2}$, $|f_{5}'|=9.8\times10^{-2}$, $|f_{5}''|=4\times10^{-2}$, $m_{N}=212$, $A_{0}=570$, $m_{E}=360$, $m_{0}=300$, $\chi_{3}=0.3$, $\chi_{3}'=0.2$, $\chi_{3}''=0.6$, $\chi_{4}=2.8$, $\chi_{4}'=0.1$, $\chi_{4}''=0.5$, $\chi_{5}=1.9$, $\chi_{5}'=0.5$ and $\chi_{5}''=0.7$. All masses are in GeV and the phases in rad.} \label{fig1}
\end{center}
\end{figure}

\cref{fig2} exhibits the variation of the neutrinos magnetic moments with  the trilinear coupling $A_{0}$ in the range $150-600$ GeV where in the analysis we assumed $A_{\tau}=A_{E}=A_{\mu}=A_{e}=A_{0}$. The trilinear coupling
appears in the slepton mass squared matrix and thus affects the chargino-slepton loop contribution. As in \cref{fig1}
the magnetic moment {analysis} corresponding to $\tan\beta =40, 50, 60$ are displayed  while the corresponding 
analysis for the $\nu_3$ lifetime are exhibited by marked lines. A sample point in the analysis is 
$\tan\beta=50$ (long-dashed curve) and $A_{0}=600$ GeV where $\mu_{2}$ takes a value $\sim1.1\times10^{-10}\mu_{B}$ and $\tau_{\nu_{3}}\sim7\times10^{13}$ yrs.\\

\begin{figure}[H]
\begin{center}
\hfill
\subfigure[The muon neutrino magnetic moment and the tau neutrino lifetime versus $A_{0}$.]
{\includegraphics[width=8cm]{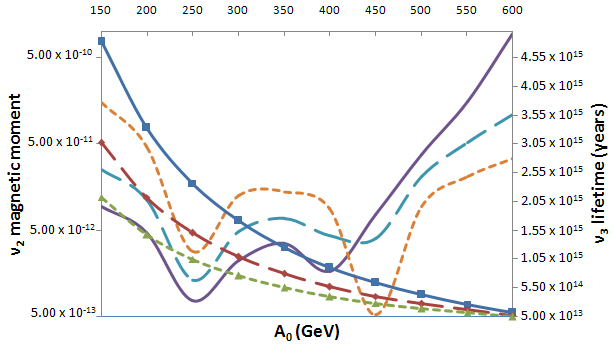}}
\hfill
\subfigure[The electron neutrino magnetic moment and the tau neutrino lifetime versus $A_{0}$.]
{\includegraphics[width=8cm]{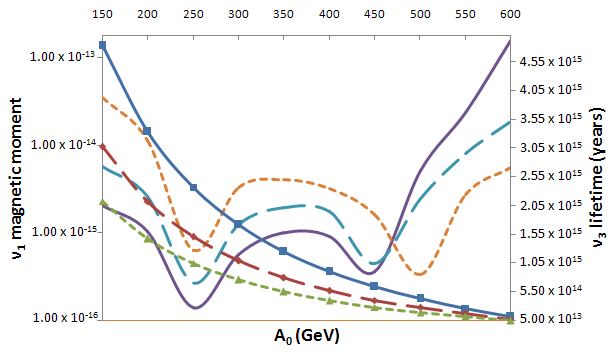}}
\hfill
\caption{A display of the neutrino magnetic moments and $\nu_{3}$ lifetime as a function of the trilinear coupling $A_{0}$ in the range 150-600 GeV. The left panel gives the $\nu_{2}$ magnetic moment and the right panel gives the $\nu_{1}$ magnetic moment along with the $\nu_{3}$ lifetime. The curves correspond to $\tan\beta=40$ (solid), $\tan\beta=50$ (long dashed), and $\tan\beta=60$ (short dashed). The marked curves correspond to the $\nu_{3}$ lifetime. Other parameters have the values $m_{2}=297.8$, $\mu=100$, $|f_{3}|=7\times10^{-8}$, $|f_{3}'|=5\times10^{-8}$, $|f_{3}''|=8\times10^{-9}$, $|f_{4}|=57$, $|f_{4}'|=42$, $|f_{4}''|=42$, $|f_{5}|=8.11\times10^{-2}$, $|f_{5}'|=9.8\times10^{-2}$, $|f_{5}''|=4\times10^{-2}$, $m_{N}=212$, $m_{E}=360$, $m_{0}=300$, $\chi_{3}=0.3$, $\chi_{3}'=0.2$, $\chi_{3}''=0.6$, $\chi_{4}=3.1$, $\chi_{4}'=0.1$, $\chi_{4}''=0.5$, $\chi_{5}=1.9$, $\chi_{5}'=0.5$ and $\chi_{5}''=0.7$. All masses are in GeV and the phases in rad.}\label{fig2}
\end{center}
\end{figure}

\cref{fig4} exhibits the variation of the neutrino magnetic moments versus the magnitude of the coupling $f_{3}$ for
$\tan\beta =40,50,60$ starting from the bottom solid curve going to the top short-dashed curve.
 As in the previous figures the unmarked curves are for the neutrinos magnetic moments
while the marked ones are for the $\nu_3$ lifetime.  The analysis of the figure shows that 
  the muon neutrino magnetic moment can take values of  the order of $10^{-10}\mu_{B}$ while the electron neutrino magnetic moment  can reaches values $\sim 2\times10^{-12}\mu_{B}$.  Also
  {as} in the previous case the constraints on $\Delta m_{31}^{2}$, $\Delta m_{21}^{2}$, and on the sum of the 
  neutrino masses must be satisfied. 
  An example of such a point is 
 $|f_{3}|=7\times10^{-8}$ GeV and $\tan\beta=60$, which gives $\mu_{2}=6.6\times10^{-10}\mu_{B}$, $\mu_{1}=1.9\times10^{-12}\mu_{B}$ and $\tau_{\nu_{3}}=2.4\times10^{14}$ yrs. It is worth noting that $\mu_{1}$ is not affected much by the change in $|f_{3}|$ as can be seen from the graph, where the curves are almost horizontal straight lines.\\

\begin{figure}[H]
\begin{center}
\hfill
\subfigure[The muon neutrino magnetic moment and tau neutrino lifetime versus $|f_{3}|$.]
{\includegraphics[width=8cm]{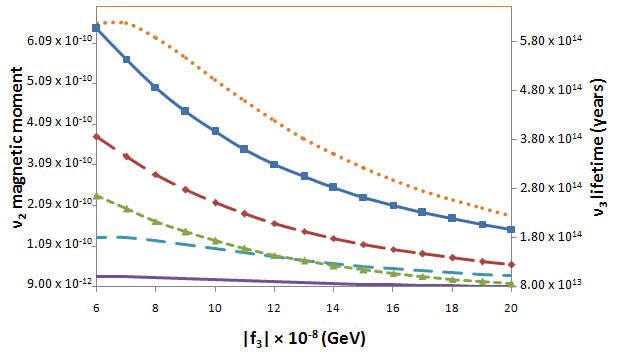}}
\hfill
\subfigure[The electron neutrino magnetic moment and the tau neutrino lifetime versus $|f_{3}|$.]
{\includegraphics[width=8cm]{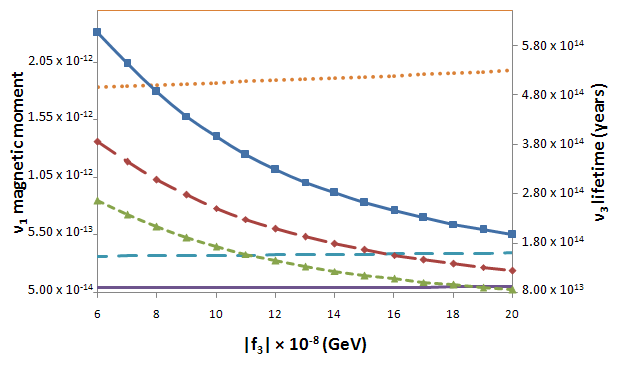}}
\hfill
\caption{A display of the neutrino magnetic moments and the $\nu_{3}$ lifetime as a function of $|f_{3}|$. The left panel gives the $\nu_{2}$ magnetic moment while the right panel gives the $\nu_{1}$ magnetic moment along with the $\nu_{3}$ lifetime. The three curves correspond to the values of $\tan\beta=40$ (solid curve), $\tan\beta=50$ (dashed), and $\tan\beta=60$ (dotted). The marked curves correspond to the $\nu_{3}$ lifetime. Other parameters have the values $m_{2}=250$, $\mu=100$, $|f_{3}'|=5\times10^{-8}$, $|f_{3}''|=8\times10^{-9}$, $|f_{4}|=55$, $|f_{4}'|=42$, $|f_{4}''|=42$, $|f_{5}|=8.11\times10^{-2}$, $|f_{5}'|=9.8\times10^{-2}$, $|f_{5}''|=4\times10^{-2}$, $m_{N}=212$, $m_{E}=360$, $m_{0}=300$, $\chi_{3}=0.3$, $\chi_{3}'=0.2$, $\chi_{3}''=0.6$, $\chi_{4}=1.8$, $\chi_{4}'=2.9$, $\chi_{4}''=0.5$, $\chi_{5}=1.9$, $\chi_{5}'=0.5$, $\chi_{5}''=0.7$ and $A_{0}=579$. All masses are in GeV and the phases in rad.} \label{fig4}
\end{center}
\end{figure}

 Next we study the effect of CP phases on the magnetic moment and on the $\nu_3$ lifetime (for a review see \cite{Ibrahim:2007fb}).
   Thus, e.g., in ~ \cite{Ibrahim:1999aj} the dependence 
 of the muon magnetic moment on CP phases was investigated and its sharp dependence on several CP phases in MSSM was found. Thus we expect that here also  the neutrino magnetic moments 
 will be sensitive to the CP phases. We begin by defining the CP phases $\chi_i$ and $\chi_i'$ (i=1-5) by
 \beq
 f_i= |f_i|e^{i\chi_i},  \  \ f_i' = |f_i'| e^{i\chi_i'}, \  \ { i=3-5} \ .
 \eeq

In \cref{fig5} we exhibit the dependence of the magnetic moments and of the $\nu_3$ lifetime  on the CP phase $\chi_4$.
The coupling $f_{4}$  appears  in the lepton mass matrix and in the slepton mass squared  matrix and thus the 
the phase enters in the W exchange contribution as well as in the chargino exchange contribution to the magnetic moment.
Variation of $\chi_4$ has no influence on the neutrino mass matrix. \cref{fig5} exhibits the variation of $\mu_1$ and $\mu_2$
and of $\nu_3$ lifetime with $\chi_4$ for three values of $m_E$, i.e., for $m_{E}=360, 460$ and $560$ GeV
starting from the solid curve down to the short-dashed curve. It is seen that significantly larger magnetic moments and significantly  {smaller}  $\nu_3$ lifetime relative to the standard model case are obtained as in the previous cases.

\begin{figure}[H]
\begin{center}
\hfill
\subfigure[The muon neutrino magnetic moment and the tau neutrino lifetime versus $\chi_{4}$.]
{\includegraphics[width=8cm]{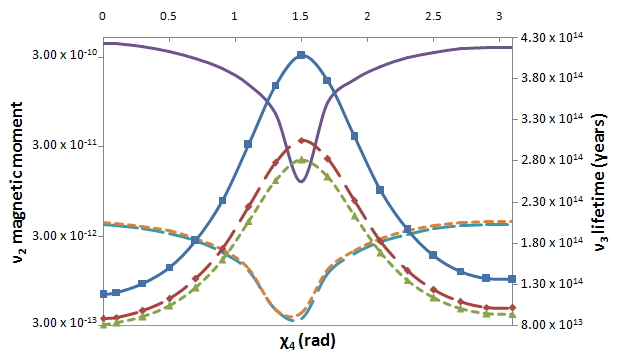}}
\hfill
\subfigure[The electron neutrino magnetic moment and the tau neutrino lifetime versus $\chi_{4}$.]
{\includegraphics[width=8cm]{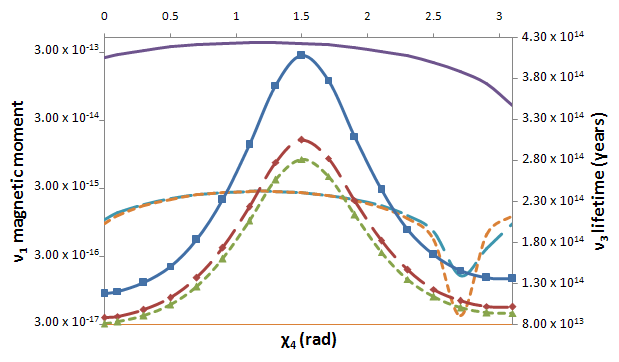}}
\hfill
\caption{A display of the neutrino magnetic moments and the $\nu_{3}$ lifetime as a function of the phase $\chi_{4}$. The left panel shows the $\nu_{2}$ magnetic moment and the right panel shows the $\nu_{1}$ magnetic moment along with the $\nu_{3}$ lifetime. The three curves correspond to the values of $m_{E}=360$ (solid curve), $m_{E}=460$ (long dashed), and $m_{E}=560$ (short dashed). The marked curves correspond to the $\nu_{3}$ lifetime. Other parameters have the values $m_{2}=250$, $\mu=100$, $|f_{3}|=7\times10^{-8}$, $|f_{3}'|=5\times10^{-8}$, $|f_{3}''|=8\times10^{-9}$, $|f_{4}|=55$, $|f_{4}'|=42$, $|f_{4}''|=42$, $|f_{5}|=8.11\times10^{-2}$, $|f_{5}'|=9.8\times10^{-2}$, $|f_{5}''|=4\times10^{-2}$, $m_{N}=212$, $m_{0}=300$, $A_{0}=579$, $\chi_{3}=0.3$, $\chi_{3}'=0.2$, $\chi_{3}''=0.6$, $\chi_{4}'=2.9$, $\chi_{4}''=0.5$, $\chi_{5}=1.9$, $\chi_{5}'=0.5$, $\chi_{5}''=0.7$ and $\tan\beta=50$. All masses are in GeV and the phases in rad.} \label{fig5}
\end{center}
\end{figure}

 In \cref{fig3} the dependence of the  neutrino magnetic moments on the phase $\chi_{5}$ is exhibited for three different values of 
 $\chi_{5}'$ from the upper solid curve down to the short-dashed curve at $\chi_{5}=0$, and $\chi_{5}'=0.5, 1.0, 1.5$ rad. 
 Here one finds that the variation of $\chi_5$ produces a dramatically different effect on $\mu_1$ vs $\mu_2$. Thus 
 $\mu_2$ has essentially a gentle dependence on $\chi_5$ while $\mu_1$ exhibits a much more rapid variation . 
Changing $\chi_{5}'$ does not alter $\mu_2$ by much which is why the curves are nearly overlapping. 
Over the entire $\chi_{5}$ range ($0\rightarrow\pi$ rad), the values of $\mu_{2}$ stretch from $\sim2.6\times10^{-10}\mu_{B}$ to $\sim4.2\times10^{-10}\mu_{B}$ which lie below the current experimental upper bounds but are tantalizingly close to it.
The corresponding $\nu_{3}$ lifetimes  are also encouraging. As for $\mu_{1}$, its value is very sensitive to variations in the phase $\chi_{5}$ and a major shift in peaks occurs for the three considered values of $\chi_{5}'$. Values in the order of $10^{-13}\mu_{B}$ are obtained in this parameter space. In the previous analysis shown in figs.~\ref{fig1} and~\ref{fig2}, the neutrino masses lie in the required range and thus the three constraints  of \cref{6.1a,6.1b,6.1c}
 were satisfied. Here, changing $\chi_{5}$ will impact the neutrino diagonalizing matrices $D_{R}^{\nu}$ and $D_{L}^{\nu}$ 
and one needs to make certain that the 
neutrino eigenmasses lie in  the acceptable range. In Fig.~\ref{fig3}, one finds that for the parameter point
$\chi_{5}=1.9$ rad and $\chi_{5}'=0.5$ {rad}
all constraints are satisfied along with desirable values of the $\nu_{3}$ lifetime, and of $\mu_{2}$ and $\mu_{1}$, i.e., here 
one has  
$\mu_{2}=3.6\times10^{-10}\mu_{B}$, $\mu_{1}=1.2\times10^{-13}\mu_{B}$ and $\nu_{3}=8.2\times10^{13}$ yrs. \\

\begin{figure}[H]
\begin{center}
\hfill
\subfigure[The muon neutrino magnetic moment and tau neutrino lifetime versus $\chi_{5}$.]
{\includegraphics[width=8cm]{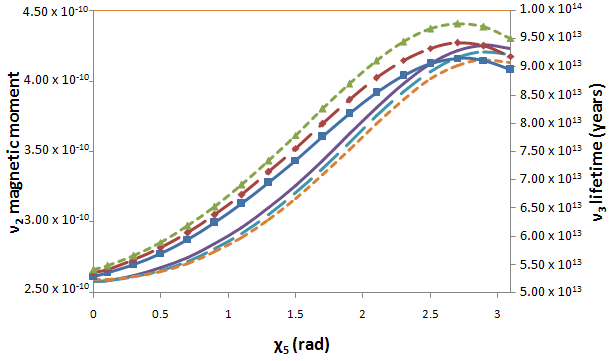}}
\hfill
\subfigure[The electron neutrino magnetic moment and tau neutrino lifetime versus $\chi_{5}$.]
{\includegraphics[width=8cm]{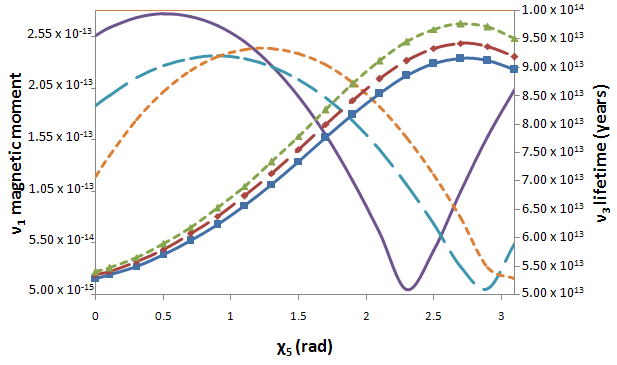}}
\hfill
\caption{A display of the neutrino magnetic moments and the $\nu_{3}$ lifetime as a function of the phase $\chi_{5}$. The left panel gives the $\nu_{2}$ magnetic moment and the right panel gives the $\nu_{1}$ magnetic moment along with the $\nu_{3}$ lifetime. The three curves correspond to  $\chi_{5}'=0.5$ (solid), $\chi_{5}'=1.0$ (long dashed), and $\chi_{5}'=1.5$ (short dashed). The marked curves correspond to the $\nu_{3}$ lifetime. Other parameters have the values $m_{2}=300$, $\mu=100$, $|f_{3}|=7\times10^{-8}$, $|f_{3}'|=5\times10^{-8}$, $|f_{3}''|=8\times10^{-9}$, $|f_{4}|=57$, $|f_{4}'|=42$, $|f_{4}''|=42$, $|f_{5}|=8.11\times10^{-2}$, $|f_{5}'|=9.8\times10^{-2}$, $|f_{5}''|=4\times10^{-2}$, $m_{N}=212$, $m_{E}=360$, $m_{0}=300$, $\chi_{3}=0.3$, $\chi_{3}'=0.2$, $\chi_{3}''=0.6$, $\chi_{4}=2.8$, $\chi_{4}'=0.1$, $\chi_{4}''=0.5$, $A_{0}=579$, $\tan\beta=60$ and $\chi_{5}''=0.7$. All masses are in GeV and the phases in rad.} \label{fig3}
\end{center}
\end{figure}

In summary, the analysis of Figs.~\ref{fig1}-~{\ref{fig3}} shows that the neutrino magnetic moments as low as the current experimental lower limits can be obtained with a vectorlike  generation. These magnetic moments are up to 7 orders of magnitude larger than with the Standard Model interactions and  within the reach of improved experiment.\\

 Finally we compare our results with some previous works where neutrino magnetic moments much larger
than given in Eq.(\ref{1.1}) were achieved. Thus  the analysis of~\cite{Gozdz:2006iz,Gozdz:2006jv}
achieved neutrino magnetic moments as large as $(10^{-17}- 10^{-16}) \mu_B$ which are up to
five to six 
 orders of magnitude larger than given by Eq.(\ref{1.1}) when one uses a generic neutrino mass of $10^{-3}$ eV
 in \cref{1.1}.
  Our predicted neutrino moments 
fall in the range $(10^{-12}-10^{-14})\mu_B$  and are up to $(10^8-10^{10})$ orders of magnitude larger than the
result of Eq.(\ref{1.1}). Further, the analysis of ~\cite{Gozdz:2006iz,Gozdz:2006jv}
is based on models with explicit R -parity violation and the low energy signatures of such models
will have no large missing energy signals which are typically associated with R-parity conserving 
supersymmetric models. In contrast our analysis is done in extensions of MSSM with R parity 
conservation. Thus the set up as well as the predictions of our work are very different from the
works of ~\cite{Gozdz:2006iz,Gozdz:2006jv}.

\section{Conclusion\label{sec7}}

It is well known that the neutrino Dirac magnetic moment computed using the standard model interactions is
far too small to be tested by experiment since it would require an increase in sensitivity
by several orders of magnitude which  appears out of reach in the current and in the  near future
experiments.  Since the neutrino magnetic moment in the standard model is too small to be observed,  an observation of a much
larger moment will be a clear indication of new physics beyond the standard model. 
 In this work we have carried out an analysis of
the neutrino magnetic moments within in an extension of MSSM which includes a vectorlike 
leptonic generation which contains a fourth leptonic generation along with
its mirrors.
{We assume terms in the superpotential of the theory that mix  the ordinary  leptonic generation with 
the vectorlike  generation.} 
The analysis of the neutrino magnetic moment within this framework
shows that a magnetic moment for the neutrinos as large as $(10^{-12}-10^{-14}) \mu_B$ can be obtained.
{These values thus lie} within  reach of improved experiment in the future.
The analysis is of importance since neutrino magnetic
moments have  implications for particle physics as well as for astrophysical phenomena,
and as already mentioned 
an observation of it of a size significantly larger than predicted with the standard model like interactions 
would provide a clear indication of new physics beyond the standard model.\\

\noindent
{\em Acknowledgments}:

This research was supported in part by the NSF Grant PHY-1314774 and 
DE-AC02-05CH11231.

\section{Appendix: Further details on the scalar mass  squared matrices   \label{sec8}}
In this Appendix  we give further details of the structure of the slepton mass matrices.
The mass terms arising from the superpotential  are given by
\beq
{\cal L}^{\rm mass}_F= {\cal L}_C^{\rm mass} +{\cal L}_N^{\rm mass}\ , 
\eeq
where  ${\cal L}_C^{\rm mass}$ gives the mass terms for the charged leptons while
$ {\cal L}_N^{mass}$ gives the mass terms for the  neutrinos. For ${\cal L}_C^{\rm mass}$ we have
\begin{gather}
-{\cal L}_C^{\rm mass} =\left(\frac{v^2_2 |f'_2|^2}{2} +|f_3|^2+|f_3'|^2+|f_3''|^2\right)\tilde E_R \tilde E^*_R
%\nonumber\\
+\left(\frac{v^2_2 |f'_2|^2}{2} +|f_4|^2+|f_4'|^2+|f_4''|^2\right)\tilde E_L \tilde E^*_L\nonumber\\
+\left(\frac{v^2_1 |f_1|^2}{2} +|f_4|^2\right)\tilde \tau_R \tilde \tau^*_R
+\left(\frac{v^2_1 |f_1|^2}{2} +|f_3|^2\right)\tilde \tau_L \tilde \tau^*_L
+\left(\frac{v^2_1 |h_1|^2}{2} +|f_4'|^2\right)\tilde \mu_R \tilde \mu^*_R\nonumber\\
+\left(\frac{v^2_1 |h_1|^2}{2} +|f_3'|^2\right)\tilde \mu_L \tilde \mu^*_L
+\left(\frac{v^2_1 |h_2|^2}{2} +|f_4''|^2\right)\tilde e_R \tilde e^*_R
+\left(\frac{v^2_1 |h_2|^2}{2} +|f_3''|^2\right)\tilde e_L \tilde e^*_L\nonumber\\
+\Bigg\{-\frac{f_1 \mu^* v_2}{\sqrt{2}} \tilde \tau_L \tilde \tau^*_R
-\frac{h_1 \mu^* v_2}{\sqrt{2}} \tilde \mu_L \tilde \mu^*_R
 -\frac{f'_2 \mu^* v_1}{\sqrt{2}} \tilde E_L \tilde E^*_R
+\left(\frac{f'_2 v_2 f^*_3}{\sqrt{2}}  +\frac{f_4 v_1 f^*_1}{\sqrt{2}}\right) \tilde E_L \tilde \tau^*_L\nonumber\\
+\left(\frac{f_4 v_2 f'^*_2}{\sqrt{2}}  +\frac{f_1 v_1 f^*_3}{\sqrt{2}}\right) \tilde E_R \tilde \tau^*_R
+\left(\frac{f'_3 v_2 f'^*_2}{\sqrt{2}}  +\frac{h_1 v_1 f'^*_4}{\sqrt{2}}\right) \tilde E_L \tilde \mu^*_L
+\left(\frac{f'_2 v_2 f'^*_4}{\sqrt{2}}  +\frac{f'_3 v_1 h^*_1}{\sqrt{2}}\right) \tilde E_R \tilde \mu^*_R\nonumber\\
+\left(\frac{f''^*_3 v_2 f'_2}{\sqrt{2}}  +\frac{f''_4 v_1 h^*_2}{\sqrt{2}}\right) \tilde E_L \tilde e^*_L
+\left(\frac{f''_4 v_2 f'^*_2}{\sqrt{2}}  +\frac{f''^*_3 v_1 h^*_2}{\sqrt{2}}\right) \tilde E_R \tilde e^*_R
+f'_3 f^*_3 \tilde \mu_L \tilde \tau^*_L +f_4 f'^*_4 \tilde \mu_R \tilde \tau^*_R\nonumber\\
+f_4 f''^*_4 \tilde {e}_R \tilde{\tau}^*_R
+f''_3 f^*_3 \tilde {e}_L \tilde{\tau}^*_L
+f''_3 f'^*_3 \tilde {e}_L \tilde{\mu}^*_L
+f'_4 f''^*_4 \tilde {e}_R \tilde{\mu}^*_R
-\frac{h_2 \mu^* v_2}{\sqrt{2}} \tilde{e}_L \tilde{e}^*_R
+H.c. \Bigg\}
\end{gather}

For ${\cal L}_N^{\rm mass}$ we have
\begin{multline}
-{\cal L}_N^{\rm mass}=
\left(\frac{v^2_1 |f_2|^2}{2}
 +|f_3|^2+|f_3'|^2+|f_3''|^2\right)\tilde N_R \tilde N^*_R\\
 +\left(\frac{v^2_1 |f_2|^2}{2}+|f_5|^2+|f_5'|^2+|f_5''|^2\right)\tilde N_L \tilde N^*_L
+\left(\frac{v^2_2 |f'_1|^2}{2}+|f_5|^2\right)\tilde \nu_{\tau R} \tilde \nu^*_{\tau R}\\
+\left(\frac{v^2_2 |f'_1|^2}{2}
+|f_3|^2\right)\tilde \nu_{\tau L} \tilde \nu^*_{\tau L}
+\left(\frac{v^2_2 |h'_1|^2}{2}
+|f_3'|^2\right)\tilde \nu_{\mu L} \tilde \nu^*_{\mu L}
+\left(\frac{v^2_2 |h'_1|^2}{2}
+|f_5'|^2\right)\tilde \nu_{\mu R} \tilde \nu^*_{\mu R}\nonumber\\
+\left(\frac{v^2_2 |h'_2|^2}{2}
+|f_3''|^2\right)\tilde \nu_{e L} \tilde \nu^*_{e L}
+\left(\frac{v^2_2 |h'_2|^2}{2}
+|f_5''|^2\right)\tilde \nu_{e R} \tilde \nu^*_{e R}\nonumber\\
+\Bigg\{ -\frac{f_2 \mu^* v_2}{\sqrt{2}} \tilde N_L \tilde N^*_R
-\frac{f'_1 \mu^* v_1}{\sqrt{2}} \tilde \nu_{\tau L} \tilde \nu^*_{\tau R}
-\frac{h'_1 \mu^* v_1}{\sqrt{2}} \tilde \nu_{\mu L} \tilde \nu^*_{\mu R}
+\left(\frac{f_5 v_2 f'^*_1}{\sqrt{2}}  -\frac{f_2 v_1 f^*_3}{\sqrt{2}}\right) \tilde N_L \tilde \nu^*_{\tau L}\nonumber\\
+\left(\frac{f_5 v_1 f^*_2}{\sqrt{2}}  -\frac{f'_1 v_2 f^*_3}{\sqrt{2}}\right) \tilde N_R \tilde \nu^*_{\tau R}
+\left(\frac{h'_1 v_2 f'^*_5}{\sqrt{2}}  -\frac{f'_3 v_1 f^*_2}{\sqrt{2}}\right) \tilde N_L \tilde \nu^*_{\mu L}
+\left(\frac{f''_5 v_1 f^*_2}{\sqrt{2}}  -\frac{f''^*_3 v_2 h'_2}{\sqrt{2}}\right) \tilde N_R \tilde \nu^*_{e R}\nonumber\\
+\left(\frac{h'^*_2 v_2 f''_5}{\sqrt{2}}  -\frac{f''^*_3 v_1 f_2}{\sqrt{2}}\right) \tilde N_L \tilde \nu^*_{e L}
+\left(\frac{f'_5 v_1 f^*_2}{\sqrt{2}}  -\frac{h'_1 v_2 f'^*_3}{\sqrt{2}}\right) \tilde N_R \tilde \nu^*_{\mu R}\nonumber\\
+f'_3 f^*_3 \tilde \nu_{\mu L} \tilde \nu_{\tau^*_L} +f_5 f'^*_5 \tilde \nu_{\mu R} \tilde \nu^*_{\tau R}
-\frac{h'_2 \mu^* v_1}{\sqrt{2}} \tilde{\nu}_{e L} \tilde{\nu}^*_{e R}\\
+f''_3 f^*_3   \tilde{\nu}_{e L} \tilde{\nu}^*_{\tau L}
+f_5 f''^*_5   \tilde{\nu}_{e R} \tilde{\nu}^*_{\tau R}
+f''_3 f'^*_3   \tilde{\nu}_{e L} \tilde{\nu}^*_{\mu L}
+f'_5 f''^*_5   \tilde{\nu}_{e R} \tilde{\nu}^*_{\mu R}
+H.c. \Bigg\}.
\label{11b}
\end{multline}

We define the scalar mass squared   matrix $M^2_{\tilde \tau}$  in the basis $(\tilde  \tau_L, \tilde E_L, \tilde \tau_R,
\tilde E_R, \tilde \mu_L, \tilde \mu_R, \tilde e_L, \tilde e_R)$. We  label the matrix  elements of these as $(M^2_{\tilde \tau})_{ij}= M^2_{ij}$ where the elements of the matrix are given by 
\begin{gather}
M^2_{11}=\tilde M^2_{\tau L} +\frac{v^2_1|f_1|^2}{2} +|f_3|^2 -m^2_Z \cos 2 \beta \left(\frac{1}{2}-\sin^2\theta_W\right), \nonumber\\
M^2_{22}=\tilde M^2_E +\frac{v^2_2|f'_2|^2}{2}+|f_4|^2 +|f'_4|^2+|f''_4|^2 +m^2_Z \cos 2 \beta \sin^2\theta_W, \nonumber\\
M^2_{33}=\tilde M^2_{\tau} +\frac{v^2_1|f_1|^2}{2} +|f_4|^2 -m^2_Z \cos 2 \beta \sin^2\theta_W, \nonumber\\
M^2_{44}=\tilde M^2_{\chi} +\frac{v^2_2|f'_2|^2}{2} +|f_3|^2 +|f'_3|^2+|f''_3|^2 +m^2_Z \cos 2 \beta \left(\frac{1}{2}-\sin^2\theta_W\right), \nonumber\\
M^2_{55}=\tilde M^2_{\mu L} +\frac{v^2_1|h_1|^2}{2} +|f'_3|^2 -m^2_Z \cos 2 \beta \left(\frac{1}{2}-\sin^2\theta_W\right), \nonumber\\
M^2_{66}=\tilde M^2_{\mu} +\frac{v^2_1|h_1|^2}{2}+|f'_4|^2 -m^2_Z \cos 2 \beta \sin^2\theta_W, \nonumber
\end{gather}

\begin{gather}
M^2_{77}=\tilde M^2_{e L} +\frac{v^2_1|h_2|^2}{2}+|f''_3|^2 -m^2_Z \cos 2 \beta \left(\frac{1}{2}-\sin^2\theta_W\right), \nonumber\\
M^2_{88}=\tilde M^2_{e} +\frac{v^2_1|h_2|^2}{2}+|f''_4|^2 -m^2_Z \cos 2 \beta \sin^2\theta_W, \nonumber\\
M^2_{12}=M^{2*}_{21}=\frac{ v_2 f'_2f^*_3}{\sqrt{2}} +\frac{ v_1 f_4 f^*_1}{\sqrt{2}} ,\nonumber\\
M^2_{13}=M^{2*}_{31}=\frac{f^*_1}{\sqrt{2}}(v_1 A^*_{\tau} -\mu v_2),\nonumber\\
M^2_{14}=M^{2*}_{41}=0, M^2_{15} =M^{2*}_{51}=f'_3 f^*_3,\nonumber\\
 M^{2*}_{16}= M^{2*}_{61}=0,  M^{2*}_{17}= M^{2*}_{71}=f''_3 f^*_3,  M^{2*}_{18}= M^{2*}_{81}=0,
M^2_{23}=M^{2*}_{32}=0,\nonumber\\
M^2_{24}=M^{2*}_{42}=\frac{f'^*_2}{\sqrt{2}}(v_2 A^*_{E} -\mu v_1),  M^2_{25} = M^{2*}_{52}= \frac{ v_2 f'_3f'^*_2}{\sqrt{2}} +\frac{ v_1 h_1 f^*_4}{\sqrt{2}} ,\nonumber\\
 M^2_{26} =M^{2*}_{62}=0,  M^2_{27} =M^{2*}_{72}=  \frac{ v_2 f''_3f'^*_2}{\sqrt{2}} +\frac{ v_1 h_1 f'^*_4}{\sqrt{2}},  M^2_{28} =M^{2*}_{82}=0, \nonumber\\
M^2_{34}=M^{2*}_{43}= \frac{ v_2 f_4 f'^*_2}{\sqrt{2}} +\frac{ v_1 f_1 f^*_3}{\sqrt{2}}, M^2_{35} =M^{2*}_{53} =0, M^2_{36} =M^{2*}_{63}=f_4 f'^*_4,\nonumber\\
 M^2_{37} =M^{2*}_{73} =0,  M^2_{38} =M^{2*}_{83} =f_4 f''^*_4,
M^2_{45}=M^{2*}_{54}=0, M^2_{46}=M^{2*}_{64}=\frac{ v_2 f'_2 f'^*_4}{\sqrt{2}} +\frac{ v_1 f'_3 h^*_1}{\sqrt{2}}, \nonumber\\
 M^2_{47} =M^{2*}_{74}=0,  M^2_{48} =M^{2*}_{84}=  \frac{ v_2 f'_2f''^*_4}{\sqrt{2}} +\frac{ v_1 f''_3 h^*_2}{\sqrt{2}},\nonumber\\
M^2_{56}=M^{2*}_{65}=\frac{h^*_1}{\sqrt{2}}(v_1 A^*_{\mu} -\mu v_2),
 M^2_{57} =M^{2*}_{75}=f''_3 f'^*_3,  M^2_{58} =M^{2*}_{85}=0,  M^2_{67} =M^{2*}_{76}=0,\nonumber\\
 M^2_{68} =M^{2*}_{86}=f'_4 f''^*_4,  M^2_{78}=M^{2*}_{87}=\frac{h^*_2}{\sqrt{2}}(v_1 A^*_{e} -\mu v_2)
\label{14}
\end{gather}

Here the terms $M^2_{11}, M^2_{13}, M^2_{31}, M^2_{33}$ arise from soft
breaking in the  sector $\tilde \tau_L, \tilde \tau_R$,
the terms $M^2_{55}, M^2_{56}, M^2_{65}, M^2_{66}$ arise from soft
breaking in the  sector $\tilde \mu_L, \tilde \mu_R$,
the terms $M^2_{77}, M^2_{78}, M^2_{87}, M^2_{88}$ arise from soft
breaking in the  sector $\tilde e_L, \tilde e_R$
 and
the terms
$M^2_{22}, M^2_{24},$  $M^2_{42}, M^2_{44}$ arise from soft
breaking in the  sector $\tilde E_L, \tilde E_R$. The other terms arise  from mixing between the staus, smuons and
the mirrors.  We assume that all the masses are of the electroweak size
so all the terms enter in the mass squared  matrix.  We diagonalize this hermitian mass squared  matrix  by the
 unitary transformation
$
 \tilde D^{\tau \dagger} M^2_{\tilde \tau} \tilde D^{\tau} = diag (M^2_{\tilde \tau_1},
M^2_{\tilde \tau_2}, M^2_{\tilde \tau_3},  M^2_{\tilde \tau_4},  M^2_{\tilde \tau_5},  M^2_{\tilde \tau_6},  M^2_{\tilde \tau_7},  M^2_{\tilde \tau_8} )$. {For a further clarification of the notation see~\cite{Ibrahim:2012ds}}.\\

\end{document}